# Role of Ionized Impurity and Interface Roughness Scatterings in the Electronic Transport of InAs/GaSb Type II Superlattices at Low Temperatures

S. Safa[1a], A Asgari[a, b]

[a]Research Institute for Applied Physics and Astronomy, University of Tabriz,

Tabriz 51665-163, Iran

[b]School of Electrical, Electronic and Computer Engineering, The University of Western Australia,

Crawley, WA 6009, Australia

**Abstract:**

The in-plane electron mobility has been calculated in InAs/GaSb type-II superlattices at low temperatures. The interface roughness scattering and ionized impurity scattering are investigated as the dominant scattering mechanisms in limiting the electron transport at low temperatures. For this purpose, the band structures and wave functions of electrons in such superlattices are calculated by solving the K.P Hamiltonian using the numerical Finite Difference method. The scattering rates have obtained for different temperatures and structural parameters. We show that the scattering rates are high in thin-layer superlattices and the mobility rises as the temperature increases in low-temperature regime.

1. Introduction

The InAs/GaSb superlattices (SLs) have recently attracted many attentions in device applications. The unique electronic properties of these systems have been make them operative in novel electronic and photonic devices such as infrared photodetectors [1] and high current density tunnel diodes [2–4]. The special broken band alignment in type II heterostructures such as InAs/GaSb leads to separation of electrons and holes in InAs and GaSb layers [5]. Since mobility is a very important parameter for semiconductor materials, almost always, higher mobility leads to a better device performance, with other things equal. This high carrier mobility can be achieved by reduction in carrier scattering in these multi-layer structures [6-11]. Such theoretical studies of diffusive carrier transport in superlattices were developed by Mori and Ando [12], Dharssi and Butcher [13], Szmulowicz et al [14, 15, 16, 17] and others [18, 19].

---
[1] Author to whom correspondence should be addressed. Electronic mail:
safa@tabrizu.ac.ir. Tel.: 98-41-33342564. Fax: 98-41-33347050.

In this literature, we have reported on the results of electron mobility calculations in type II InAs/GaSb superlattices by considering the ionized impurity scattering and interface roughness scattering which are the dominant scattering mechanisms in these systems at low temperatures. Although, previously these scattering mechanisms were studied separately in some specified superlattices [17, 20, 21], but a plenary view is needed in order to illustrate the low temperature scattering mechanisms in relation with wells and barriers thicknesses in a SL. For this purpose, an accurate way is needed to calculate the electronic band structure and wave functions which can be achieved through numerical Finite Difference method for a K.P Hamiltonian. In these calculations, the precise values of each SL layer have been included in the matrix range, so the exact wave functions for interfaces and their energies are determined. Then the electron mobility has been calculated, and the effects of temperature and SL structural parameters on the mobility have been analyzed. This means that we have presented a model through a computational algorithm, which calculates the electron mobility with respect to arbitrarily specified values of structural and interface parameters, ionized impurity scattering and the operating temperature. Also we have compared our calculated results with published experimental data, which shows a good agreement.

## 2. Analytical Formalism

### 2.1. Electronic structure

A theoretical eight-band **k.p** finite difference method is developed and applied to calculate the electronic band structures and wave function of the InAs/GaSb SLs. The basic idea of this method is to use a layer-orbital basis to express the Hamiltonian as accurately as possible:

$$\left( H_0 + \frac{\hbar}{m_0} K.P + \frac{\hbar^2 k^2}{2m_0} + \frac{\hbar}{4m_0^2 c^2} (\sigma \times \nabla V) \cdot (\hbar K + P) \right) \psi_{n,k} = E_{n,k} \psi_{n,k} \qquad (1)$$

where

$$H_0 = \frac{p^2}{2m_0} + V(z) \tag{2}$$

The full Hamiltonian will generate the matrix elements $H_{mn} = \langle u_{m,0} | H | u_{n,0} \rangle$. The general wave function is a linear expansion of basis functions $u_{n,0} \uparrow$ and $u_{n,0} \downarrow$, $n \in \{s, p_x, p_y, p_z\}$, as

$$\psi_{n,k} = \sum_m F_{n,m}\uparrow u_{m,0}\uparrow + \sum_m F_{n,m}\downarrow u_{m,0}\downarrow . \tag{3}$$

Now we replace $k_z$ in the Hamiltonian by the differential operator $-i\frac{\partial}{\partial z}$ and the Finite Difference method is used to express the derivatives as

$$\left.\frac{\partial F}{\partial z}\right|_{z=z(l)} = \frac{F_{i+1} - F_{i-1}}{2h} \quad \& \quad \left.\frac{\partial^2 F}{\partial z^2}\right|_{z=z(l)} = \frac{F_{i+1} - 2F_i + F_{i-1}}{h^2} \tag{4}$$

, where $h$ is the distance between two adjacent dia-atomic layers, i.e. $h=a/2$, $a$ being the lattice constant of a unit cell. Inserting the profile values for each individual mono layer of a SL consisted of 300 InAs/GaSb periods, gives the required K.P Finite Difference matrix for the system. Solving this huge matrix results in band structure and wave functions of electron in different SLs. [22, 23, 24]

### 2.2. Scattering mechanisms

#### 2.2.1. Interface Roughness Scattering

The interface roughness scattering is the dominant scattering mechanism in SLs at low temperatures. In this section, the electron mobility is calculated due to this scattering mechanism at low temperatures by employing a relatively thin periodic structure. The interface roughness can be represented by a randomness of the interface, as described by a correlation function on the in-plane position $(r_\parallel)$, which is usually taken to have a Gaussian distribution with a characteristic roughness height of $\Delta$, and a correlation length $\Lambda$. This represents a length scale for roughness fluctuations along the interface, such that [25]

$$\langle \Delta(r_{\parallel})\Delta(r'_{\parallel})\rangle = \Delta^2 e^{-\frac{|r_{\parallel}-r'_{\parallel}|^2}{\Lambda^2}} \qquad (5)$$

and the associated perturbation potential caused by interface roughness has the form

$$\delta V(r_{\parallel},z) = V_e \Delta(r_{\parallel})[\delta(z+a)-\delta(z-a)] \qquad (6)$$

According to Fermi's golden rule, the scattering rate is given by [26]

$$\Gamma(\mathbf{k},\mathbf{k}') = \frac{2\pi}{\hbar}|M(\mathbf{k},\mathbf{k}')|^2 \delta(E(\mathbf{k})-E(\mathbf{k}')) \qquad (7)$$

,where

$$|M(\mathbf{k},\mathbf{k}')|^2 = |\langle \mathbf{k}_{\parallel},k_z|\delta V(r_{\parallel})|\mathbf{k}'_{\parallel},k'_z\rangle|^2. \qquad (8)$$

Next, from the Boltzmann equation and relaxation time approximation we obtain the relaxation time of electrons in this scattering as,

$$\frac{1}{\tau_{IR}(\mathbf{k})} = \frac{V}{(2\pi)^3}\int d\mathbf{k}'\, \Gamma(\mathbf{k},\mathbf{k}')\frac{(1-f_0(\mathbf{k}'))}{(1-f_0(\mathbf{k}))}\times\left[1-\frac{f_0(\mathbf{k})(1-f_0(\mathbf{k}))}{f_0(\mathbf{k}')(1-f_0(\mathbf{k}'))}\frac{g(\mathbf{k}')}{g(\mathbf{k})}\right] \qquad (9)$$

, where $f_0$ is the equilibrium Fermi-Dirac distribution function and $g$ is the deviation from equilibrium in presence of electric field F:

$$g(\mathbf{k}) = \tau(\mathbf{k})\frac{e}{\hbar}\left(F\cdot\frac{\partial E}{\partial \mathbf{k}}\right)\frac{\partial f_0}{\partial E} \qquad (10)$$

*2.2.2. Ionized impurity Scattering*

There are two standard treatments in impurity scattering transport phenomena reported by Conwell and Weisskopf [27], and by Brooks and Herring [28]. Conwell and Weisskopf have derived the relaxation time for ionized impurity scattering by adopting Rutherford scattering. The latter is the method which we have used in this literature. In a semiconductor with impurity density $n_I$ and electron velocity $v(\mathbf{k}) = \frac{1}{\hbar}\frac{\partial E}{\partial \mathbf{k}}$, the relaxation time of an electron, $\tau_I$, is given by $\frac{1}{\tau_I} = n_I v A$. The differential cross-section $\sigma(\theta,\varphi)$ is defined by the probability of scattering into a small solid angle

$d\Omega = \sin\theta\, d\theta\, d\varphi$. In the case of isotropic scattering the cross section is given by $A = \int \sigma(\theta,\varphi) \sin\theta\, d\theta\, d\varphi$. Assuming the scattering to be elastic and by defining $\theta$ as the angle between the electron wave vectors $k$ and $k'$, we obtain

$$\frac{1}{\tau_I} = n_I v \int \sigma(\theta,\varphi)(1-\cos\theta)\sin\theta\, d\theta\, d\varphi \qquad (11)$$

Neglecting the screening effect within the conduction electrons, the potential induced by an ionized impurity is given by Rutherford scattering cross section [29]

$$\sigma(\theta) = \frac{1}{4} R^2 \cosec^4\left(\frac{\theta}{2}\right), \quad R = \frac{ze^2}{4\pi\varepsilon m^* v^2}. \qquad (12)$$

Considering the impurity density $n_I = N_I/L^3$ in which $N_I$ is the total concentration of ionized donors and acceptors, the scattering rate becomes:

$$\frac{1}{\tau_I(k)} = \frac{2\pi}{4L^3 \hbar}\int d\mathbf{k}'\, d\theta\, N_I \left(\frac{\partial E}{\partial \mathbf{k}'}\right)\left(\frac{ze^2\hbar}{4\pi\varepsilon m^*\left(\frac{\partial E}{\partial \mathbf{k}'}\right)^2}\right)^2 \cosec^4\left(\frac{\theta}{2}\right)(1-\cos\theta)\delta(E_{\mathbf{k}'}-E_{\mathbf{k}}) \qquad (13)$$

Eventually, one may sum over these two scattering effects to calculate the total electron relaxation time via Matheson's rule:

$$\frac{1}{\tau_{Total}(\mathbf{k})} = \frac{1}{\tau_{IR}(\mathbf{k})} + \frac{1}{\tau_I(\mathbf{k})} \qquad (14)$$

Thus the total mobility for an arbitrary superlattice structure emerges in terms of the relaxation time as

$$\mu = -e \frac{\int \tau_{Total}(\mathbf{k})\left(\frac{\partial E}{\partial \mathbf{k}}\right)^2 \left(\frac{\partial f_0}{\partial E}\right)^2 d\mathbf{k}}{\int f_0(\mathbf{k})\, d\mathbf{k}} \qquad (15)$$

## 3. Results and Discussion

An understanding of the temperature dependence of the electron mobility due to the interface roughness scattering of the InAs/GaSb systems is shown in Fig. 1 for different structural parameters. Although, the interface roughness scattering is temperature independent, but the carrier distribution function is temperature dependent, so the electron mobility via interface roughness scattering will be temperature dependent slightly, as depicted in equation (15). As shown on Fig.1-a, the mobility increases by increasing the width of the InAs wells, which is consistent with the Gold model [30]. Hence, the mobility is more sensitive to InAs-layer width fluctuations [16].

Fig.1-b shows the behavior of the electron mobility as a function of GaSb barrier thickness, possessing an optimum value for the thickness of 4*nm*. This extremum happens when the penetration of the electron wave function is balanced with the magnitude of the effective mass in such a SL.

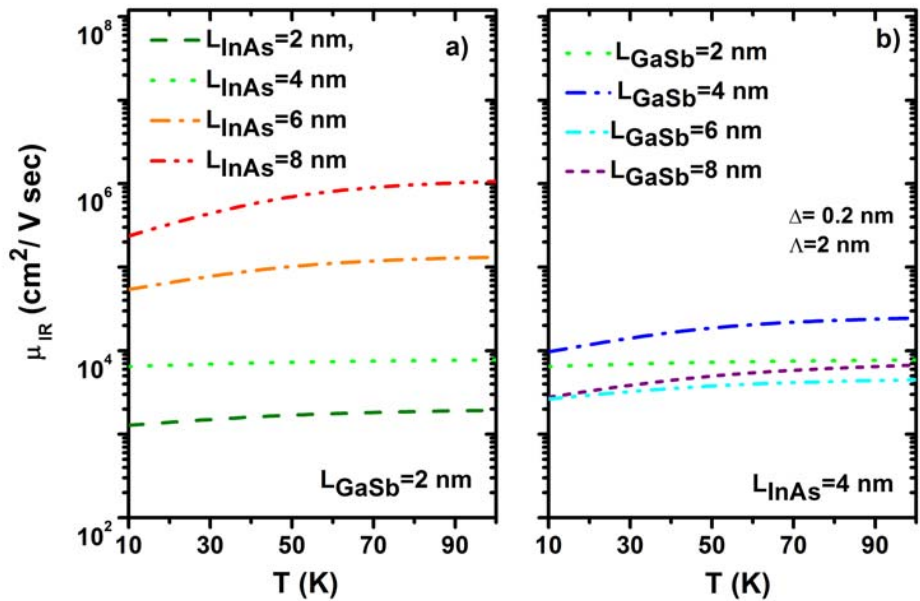

Fig.1. Electron mobility due to interface roughness scattering vs. temperature in InAs/GaSb SLs with different (a) InAs and (b) GaSb thickness.

Figure 2 shows the dependence of electron mobility due to ionized impurity scattering as a function of temperature for SLs with different InAs (a) and GaSb (b) thicknesses. The electron mobility increases rapidly by the temperature enhancement and this is because the value of $\frac{\partial f_0}{\partial \mathbf{E}}$ in equation (15), is an ascending function of temperature. Since the band structure is calculated in a constant temperature, the only temperature dependence of the mobility comes from the carrier distribution function.

The dependence of this mobility on InAs and GaSb is almost the same. According to theory, in thin layer SLs we see larger scattering rates and a reduction in electron mobilities, which is a result of the low electron velocity due to the low band slope $\left(\frac{\partial E}{\partial k}\right)$. Also, the superlattices with thin layers have a higher electron effective mass and a larger wave function leakage which are the reasons why the ionized impurity scattering rates increase. As the thickness of the layers increases, the semiconductor-semimetal transition takes place in type II SLs which causes a rise in mobilities.

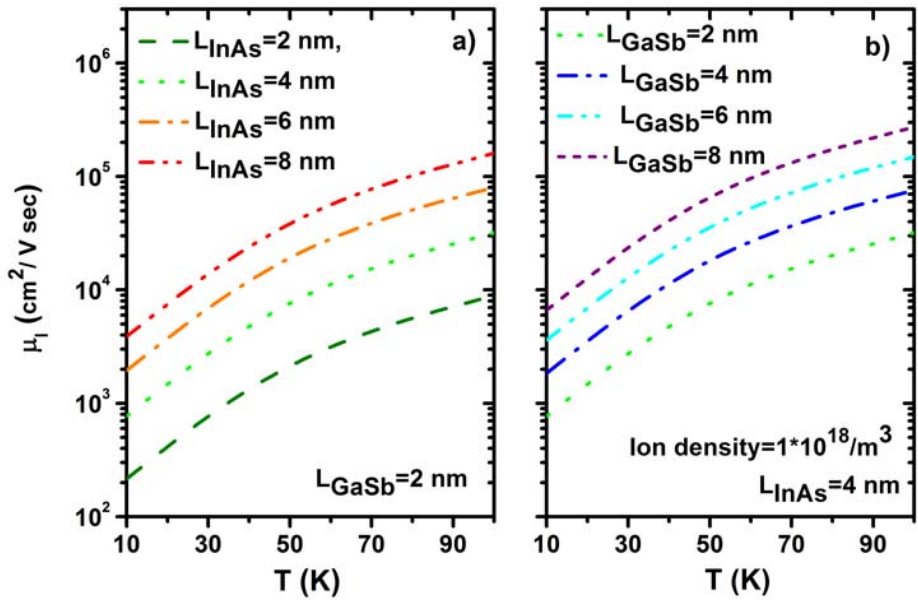

Fig.2. Electron mobility due to ionized impurity scattering vs. temperature in InAs/GaSb SLs with different (a) InAs, and (b) GaSb thickness.

The total electron mobility limited by both scattering mechanisms has been calculated via Matheson's rule as a function of temperature for different InAs/GaSb SL structures with various InAs and GaSb thicknesses as shown in Fig. 3. The behavior of the total mobility shows that at very low temperatures it is limited by ionized impurity scattering, and it is restricted by interface roughness scattering above 50 K.

The total electron mobility for InAs/GaSb SLs structures with various InAs is shown in Fig. 3-a. It has been shown that generally the mobility increases fast by increasing the InAs thickness [30]. It is evident from the figure that the mobility has an optimum at around 70 K for all structures.

In Fig. 3-b, the total electron mobility for InAs/GaSb SLs structures with various GaSb layer is depicted. The behavior of the mobility as a function of GaSb thickness is complicated. This behavior can be explained by detailed different scattering mechanisms. Furthermore, the 4 *nm*/ 4 *nm* SL has the highest mobility among the other sizes, and it is because of the lower interface roughness scattering in this system.

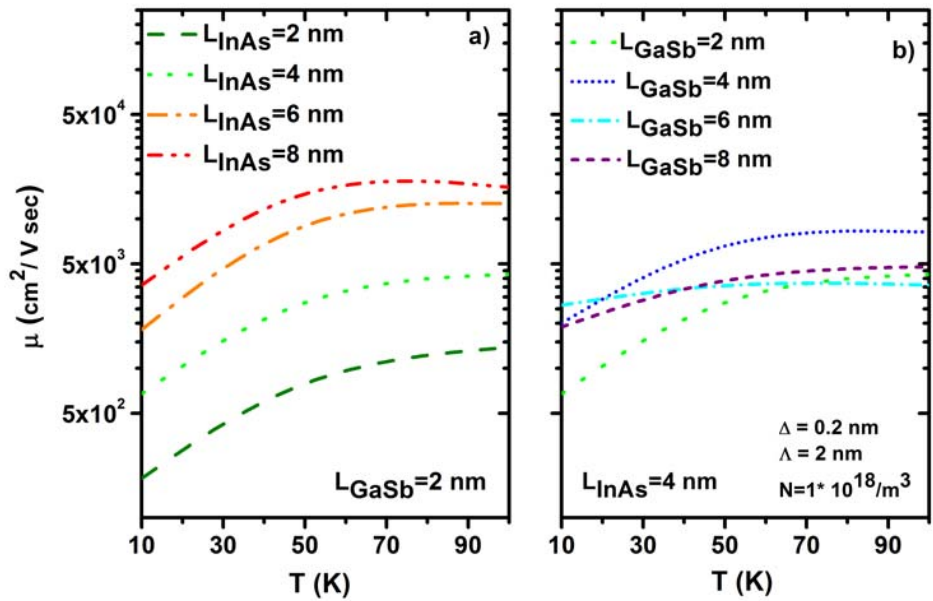

Fig.3. Total electron mobility vs. temperature in InAs/GaSb SLs with different (a) InAs, and (b) GaSb thickness.

We have compared our theoretical results with 2.7 *nm* /2.7 *nm* (InAs/GaSb) SLs which have been studied experimentally in Ref [19], and the analogy is shown in Fig. 4. By obtaining the band structure information from Finite Difference K.P algorithm for this specified superlattice and inserting the structural parameters as $\Delta = 0.1$ nm, $\Lambda = 0.5$ nm, $N_I = 5 \times 10^{15}\ m^{-3}$, one can find the total electron mobility of both scattering mechanisms of the system. The theoretical data fits very well to the experimentally measured mobilities.

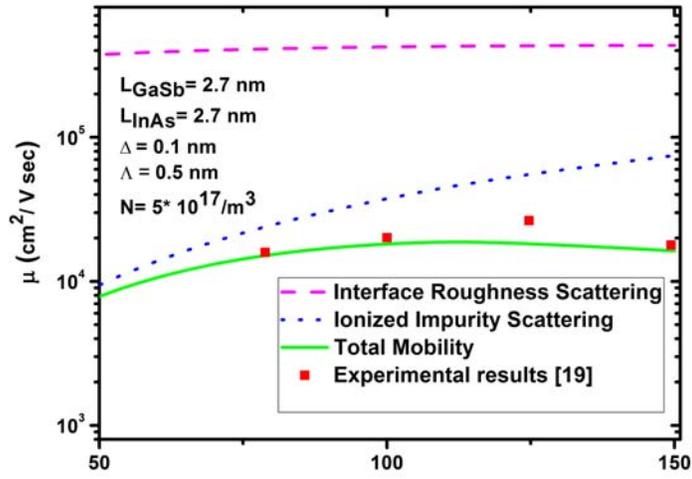

Fig.4. Total electron mobility vs. temperature in InAs/GaSb SLs with 2.7 nm InAs, and 2.7 nm GaSb. Experimental data are taken from Ref [19].

We have also studied two other InAs/GaSb superlattice structures with GaSb thickness of 2.1 nm and InAs thickness of 3.9 nm [31]. In Ref. [31] the experimental data indicates that the sample with growth at $425^0$C (Sample i) (Fig.5-a) has a higher electron mobility than the samples with growth at higher temperatures, (Sample ii) (Fig.8-b). A good agreement is seen which confirms our numerical model. This is because our calculations show that the mobility of sample (ii) is mostly limited by scattering mechanisms which belong to larger structural defects such as ionized impurity and interface roughness caused by growth at higher temperatures. The calculation shows that the ionized impurity is the most dominant scattering mechanisms at lower temperatures, although the interface roughness scattering is the source of the electron transport limitation near 100 K. Sample (i) which is a superlattice growth at low temperature has less structural defects in comparison with sample (ii). The comparison of Fig. 5-a and Fig.5-b shows that structural defect parameters such as roughness height $\Delta$, correlation length $\Lambda$ and ionized impurity density are higher in sample (ii) rather than the sample (i).

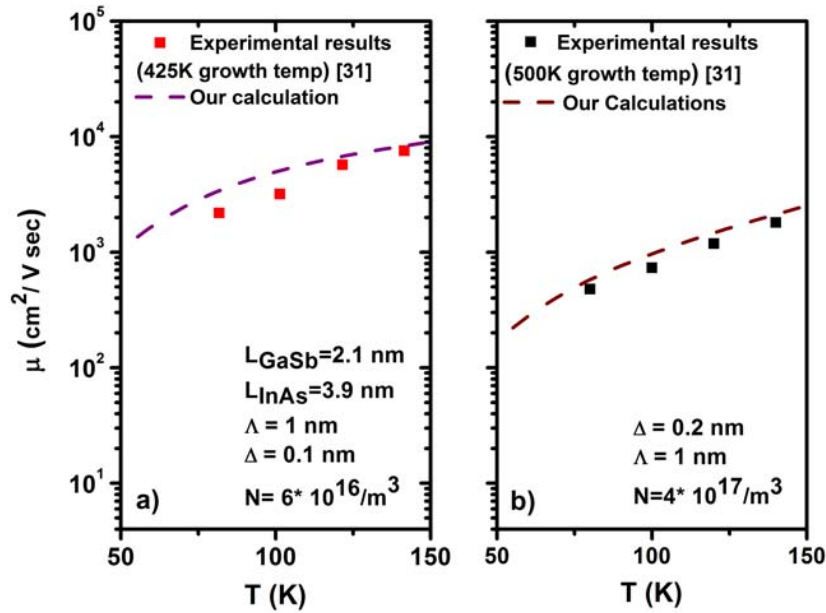

Fig.5. Electron mobility vs. temperature for the high temperature growth sample. Experimental data are taken from Ref [31].

## 4. Conclusions

We have presented a model through a computational algorithm, which calculates the electron mobility with respect to arbitrarily specified values of structural sizes, interface parameters, ionized impurity density and the operating temperature. The electron mobility in InAs/GaSb SLs were calculated by solving the corresponding Boltzmann equations using the energy spectra and wave functions obtained from the solution of the K.P Hamiltonians by the Finite Difference method. The behavior of the mobilities was examined as a function of SL parameters and was explained in terms of the underlying physics. At very low temperatures, the ionized impurity scattering is the dominant scattering mechanism and limits the mobilities. As the temperature raise, the interface roughness scattering limits the electron transports. The calculated results show that the thick well superlattices are suitable for low temperature usage. The present results reveal that mobilities have an optimum value at about 70 K., and it is the temperature in which different scattering mechanisms have the beneficial interference. Comparisons of our calculated results with published experimental data are shown to be in good agreement.